# Cooperative Energy Scheduling of Multi-Microgrids Based on Risk-Sensitive Reinforcement Learning

Rongxiang Zhang [a], Bo Li [a,b,*], Jinghua Li [a], Yuguang Song [b], Ziqing Zhu [c], Wentao Yang [a], Zhengmao Li [d], Edris Pouresmaeil [d], Joshua Y. Kim [e]

[a] School of Electrical Engineering, Guangxi University, Nanning 530004, Guangxi, China (e-mail: 2412391083@st.gxu.edu.cn; boli@gxu.edu.cn; lijinghua@gxu.edu.cn; wentaoyang@zju.edu.cn ).

[b] Department of Electrical Engineering, Tsinghua University, Beijing 100084, China (e-mail: songyg@bjtu.edu.cn ).

[c] Department of Electrical and Electronics Engineering, The Hong Kong Polytechnic University, Hong Kong (e-mail: ziqing-yancy.zhu@polyu.edu.hk ).

[d] Department of Electrical Engineering and Automation, Aalto University, Espoo 02150, Finland (e-mail: zhengmao.li@aalto.fi; edris.pouresmaeil@aalto.fi ).

[e] Department of Engineering, Charleston Southern University, North Charleston, SC 29406 USA (e-mail: ykim@csuniv.edu ).

*Abstract*—With the rapid development of distributed renewable energy, multi-microgrids play an increasingly important role in improving the flexibility and reliability of energy supply. Reinforcement learning has shown great potential in coordination strategies due to its model-free nature. Current methods lack explicit quantification of the relationship between individual and joint risk values, resulting in obscured credit assignment. Moreover, they often depend on explicit communication, which becomes inefficient as system complexity grows. To address these challenges, this paper proposes a risk-sensitive reinforcement learning framework with shared memory (RRL-SM) for multi-microgrid scheduling. Specifically, a risk-sensitive value factorization scheme is proposed to quantify the relationship between individual and joint risk values by leveraging distributional modeling and attention-based representations, thereby aligning local decisions with global risk objectives. An implicit shared-memory coordination mechanism is implemented through a global memory space to enhance the overall efficiency of decentralized decision-making. Collectively, the integrated approach delivers more reliable cooperative scheduling under renewable energy uncertainty. Simulation results show that RRL-SM reduces load-shedding risk by 84.5%, demonstrating a favorable balance between reliability and economic performance.

**Nomenclature**

**Indices and Sets**

| | |
|---|---|
| $i \in \Omega^{\mathcal{N}}$ | Index and set of microgrid agents |
| $t \in \Omega^{T}$ | Index and set of time steps |
| $b \in \Omega^{\mathrm{PV}}$ | Index and set of PV output scenarios |
| $d \in \Omega^{\mathrm{Load}}$ | Index and set of load scenarios |

**Parameters**

| | |
|---|---|
| $\Delta t$ | Time resolution |
| $T$ | Episode length |
| $\eta^{\mathrm{c}} / \eta^{\mathrm{d}}$ | Charging/discharging efficiency of energy storage system (ESS) |
| $B / D$ | Total number of PV output/load scenarios |
| $\lambda_i^{\mathrm{MT}}$ | The unit generation cost of microturbine (MT) |
| $\lambda_t^{\mathrm{in}} / \lambda_t^{\mathrm{out}}$ | Price of power purchased from/sold to the grid |
| $\lambda_i^{\mathrm{TL}}$ | Penalty cost per unit of load shedding |
| $P_{i,\max}^{\mathrm{PV}} / P_{i,\min}^{\mathrm{PV}}$ | Power output limitation of PV |
| $P_{i,\max}^{\mathrm{MT}} / P_{i,\min}^{\mathrm{MT}}$ | Power output limitation of MT |
| $R_i^{\mathrm{up}} / R_i^{\mathrm{down}}$ | Up/down ramping limit of MT |
| $P_{i,\max}^{\mathrm{c}} / P_{i,\max}^{\mathrm{d}}$ | Maximum charging/discharging power of the ESS |
| $E_{i,\max}$ | Maximum energy capacity of ESS |
| $P_{\max}^{\mathrm{GB}} / P_{\max}^{\mathrm{GS}}$ | Maximum power purchased from/sold to the grid |
| $p_b / q_d$ | Probability of PV output/load scenarios |
| $P_{i,t}^{\mathrm{PV}}$ | PV output |
| $P_{i,t}^{\mathrm{L}}$ | Load demand |
| $\mu_{b,t}$ | PV output under scenario $b$ |
| $\mu_{d,t}$ | Load demand under scenario $d$ |
| $\alpha$ | Confidence level of the risk measure |
| $\sigma$ | Penalty cost coefficient for load shedding |

**Variables**

| | |
|---|---|
| $P_{i,t}^{\mathrm{MT}}$ | MT output |
| $P_{i,t}^{\mathrm{ES}}$ | Charging and discharging power of ESS |
| $P_{i,t}^{\mathrm{c}} / P_{i,t}^{\mathrm{d}}$ | Charging/discharging power of ESS |
| $P_{i,t}^{\mathrm{TL}}$ | Load shedding |
| $SOC_{i,t}$ | State of charge (SOC) of ESS |
| $P_{i,t}^{\mathrm{GB}} / P_{i,t}^{\mathrm{GS}}$ | Power purchased from/sold to the grid |
| $\vartheta_{b,d}$ | Total load shedding loss under scenario $b,d$ |

∗ Corresponding author.
*E-mail address:* boli@gxu.edu.cn (B. Li)

# I. INTRODUCTION

Microgrids (MGs) are regarded as a key solution for integrating distributed renewable energy into distribution networks [1]. As neighboring MGs exhibit differences in load profiles and energy resources, they possess significant complementary potential. Using this potential through coordinated energy management in multi-MG systems presents an effective approach to achieving efficient supply–demand balance [2]. Practical implementations have been reported in regions such as the United States, Europe, and China. Recent studies present concrete examples, including networked community microgrids in Puerto Rico, a laboratory-scale DC multi-MG experimental platform, interconnected commercial-building multi-MG economic operation, and cooperative scheduling enabled by shared energy storage systems [3]. Despite the benefits of coordinated scheduling, the operation of multi-MG systems is subject to uncertain risks, resulting from real-time power imbalance between generation and load demand [4,5]. Such risks pose challenges to the cooperative energy scheduling of multi-MG systems.

In recent years, multi-agent reinforcement learning (MARL) has gained increasing attention as a model-free solution for coordinated energy scheduling. A challenge in cooperative MARL is ensuring that the independent decisions of individual agents lead to a coherent, globally optimal outcome. A trust region model for multi-agent action control is introduced to resolve action conflicts within the decision-making time series [6]. Value factorization methods, such as QMIX and Value-Decomposition Networks (VDN), have become a mainstream approach in cooperative MARL [7]. These methods do not account for the risks associated with uncertainty. Because value factorization is predicated on the additivity of expected returns, it is difficult to extend it to risk calculations, which are non-additive and based on return distributions.

The operational risks in multi-MG energy scheduling have motivated the development of various risk-sensitive approaches. In [8], a risk-sensitive trust region policy optimization algorithm is introduced to resolve objective conflicts in distributed decision-making through stochastic sequential decisions. In [9], a risk-based scheduling approach is proposed to minimize cost, risk, and $CO_2$ emissions in P2X-integrated multi-energy microgrids. In [10], a risk-constrained bi-level energy management strategy is proposed to optimize multi-MG scheduling by accommodating demand uncertainties. In [11], a risk-averse control framework for mobile battery energy storage systems is constructed based on hybrid risk estimation, achieving a balance between arbitrage profits and risk. However, these approaches lack explicit quantification of the relationship between individual and joint risk profiles, resulting in obscure credit assignment within complex risk-averse tasks. This ambiguity could lead to suboptimal local decision-making, as agents cannot accurately attribute global risks to their own actions.

In the MARL, existing coordination strategies often rely on explicit communication to achieve cooperation among agents. These methods typically fall into a decentralized setting with networked agents, which allows agents to share local information with each other [12]. In [13], a graph neural network-based MARL algorithm is proposed to train the distributed policies in an effective and centralized fashion. In [14], a highly-scalable global communication mechanism is proposed to enable agents with very limited fields of view to collaboratively solve complex multi-agent path finding tasks. Designing such explicit communication mechanisms is challenging. As the number of agents and problem dimensionality increase, communication overhead and network latency rise sharply, rendering the protocols inefficient. Another approach to alleviating partial observability is to enhance agent memory. In [15], a transformer-based individual working memory is employed to process an agent's self-observed factored environmental entities and its private memory. In [16], segment-level recurrence is achieved in a recurrent memory transformer by adding special memory tokens to the input or output sequence. These methods focus on optimizing individual decision-making rather than directly addressing efficient coordination among agents. Inspired by Global Workspace Theory, a shared memory mechanism replaces point-to-point communication with a shared data space for agent coordination, thereby enabling scalable decentralized decision-making [17]. However, its application to the critical domain of multi-MG energy scheduling remains a significant and unexplored research gap.

To address these limitations, this paper proposes a risk-sensitive reinforcement learning framework with shared memory (RRL-SM) for multi-microgrid scheduling. Based on our previous studies [8,18], the framework integrates two core mechanisms. On the one hand, this work pioneers a risk-sensitive value factorization within the policy gradient framework. It first establishes a risk consistency mapping between individual and joint return distributions. And then multi-head attention is employed to capture the dynamic correlations. Finally, an enhanced value estimation baseline ensures policy gradient alignment, which fosters local decisions consistent with global risk objectives. On the other hand, a shared memory mechanism distills and aggregates local information into a global memory space, allowing agents to interact implicitly. This implicit information exchange improves inter-agent coordination and decision-making efficiency.

The contributions of the work are as follows:

(1) A risk-sensitive centralized training with decentralized execution (CTDE) framework is developed. This framework establishes a new paradigm for coordinating risk-averse agents through an innovative value factorization approach, thereby achieving highly efficient risk-sensitive cooperative scheduling.

(2) A mixing network based on a multi-head attention mechanism is proposed to implement the risk-sensitive value factorization. Since each agent's contribution to the global risk varies with the system state, this attention-based mixer dynamically infer complex inter-agent dependencies and learn a state-dependent mapping to the joint risk value.

(3) A shared memory coordination mechanism is proposed. It establishes a global memory space to aggregate local observations, thereby enhancing real-time coordination capabilities and decision-making efficiency among agents.

The rest of this paper is organized as follows. Section II formulates the multi-MG scheduling model. Section III details the methodological basis for risk-sensitive MARL. Section IV describes the proposed RRL-SM framework and its network design. Section V provides the case study results, and Section VI concludes the paper.

## II. MODEL FORMULATION

### A. System Architecture

The architecture of a multi-MG system is shown in Fig. 1. This work considers multiple geographically proximate MG parks. Each MG includes microturbine (MT), photovoltaic (PV), energy storage system (ESS), and residential load. Each MG has an individual energy management system to achieve optimal scheduling. Assuming no power trading occurs between MGs, each MG trades directly with the upstream grid. Under this structure, each agent dynamically coordinates its ESS charging/discharging, MT output, and grid power trading in response to real-time supply-demand fluctuations.

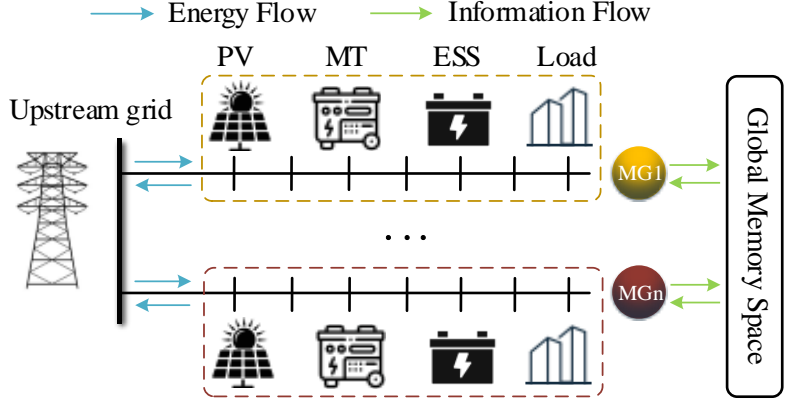

Fig. 1. The architecture of a multi-MG system.

MGs are assumed to share a common objective of minimizing operational costs while meeting load demand and maintaining secure operation. The market interaction between each MG and the upstream grid is modeled as a real-time energy purchase and sale mechanism. Each MG operates as a prosumer: it purchases power from the upstream grid at a real-time price during local supply shortages and sells its surplus power at a feed-in tariff during local oversupply. When power balance cannot be maintained due to limited line capacity or grid accommodation capability, load shedding could occur, incurring penalty costs.

### B. Modeling of Multi-MG Energy Scheduling

The objective of the multi-MG system is to minimize its total operational cost over the entire scheduling horizon. This objective is formulated as follows:

$$\min C = \sum_{t=1}^{T} \sum_{i=1}^{N} \left[ \lambda_i^{MT} P_{i,t}^{MT} + \lambda_t^{in} P_{i,t}^{GB} - \lambda_t^{out} P_{i,t}^{GS} + \lambda_i^{TL} P_{i,t}^{TL} \right] \Delta t \tag{1}$$

where $C$ denotes the total operational cost. This cost comprises four components. $\lambda_i^{MT} P_{i,t}^{MT}$ represents the operational cost of the MT. $\lambda_t^{in} P_{i,t}^{GB}$ represents the cost of purchasing power from the upstream grid. $\lambda_t^{out} P_{i,t}^{GS}$ represents the revenue from selling power to the upstream grid. $\lambda_i^{TL} P_{i,t}^{TL}$ represents the cost for load shedding.

In this study, the risk of load shedding stems primarily from the transmission capacity limit of the grid connection lines between the microgrids and the upstream grid [8]. When the total real-time power deficit of MGs (e.g., caused by a sudden drop in PV output or a surge in load) exceeds this physical limit, the system is forced to shed load as it cannot secure sufficient external support. The penalty cost for load shedding defined in the model is intended to represent the comprehensive operational losses caused by power interruption, such as loss of electricity revenue and degradation of service reliability, rather than direct economic compensation to end-users. Setting this penalty is designed to effectively guide the agents to prioritize power supply reliability.

The constraints for the Multi-MG energy scheduling problem include power balance constraints, MT output constraints, ESS operation constraints, and MG-grid trading constraints.

$$P_{i,t}^{PV} + P_{i,t}^{MT} + P_{i,t}^{d} - P_{i,t}^{c} + P_{i,t}^{GB} - P_{i,t}^{GS} = P_{i,t}^{L} - P_{i,t}^{TL} \tag{2}$$

$$P_{i,\min}^{PV} \le P_{i,t}^{PV} \le P_{i,\max}^{PV} \tag{3}$$

$$P_{i,\min}^{MT} \le P_{i,t}^{MT} \le P_{i,\max}^{MT} \tag{4}$$

$$-R_i^{down} \le P_{i,t}^{MT} - P_{i,t-1}^{MT} \le R_i^{up} \tag{5}$$

$$0 \le P_{i,t}^{c} \le P_{i,\max}^{c} \tag{6}$$

$$0 \le P_{i,t}^{d} \le P_{i,\max}^{d} \tag{7}$$

$$SOC_{i,\min} \le SOC_{i,t} \le SOC_{i,\max} \tag{8}$$

$$SOC_{i,t} = SOC_{i,t-1} + \frac{(P_{i,t}^c \times \eta^c - P_{i,t}^d / \eta^d)}{E_{i,\max}} \tag{9}$$

$$P_{i,t}^{GB} \times P_{i,t}^{GS} = 0 \tag{10}$$

$$0 \leq \sum_{i=1}^{N} P_{i,t}^{GB} \leq P_{\max}^{GB} \tag{11}$$

$$0 \leq \sum_{i=1}^{N} P_{i,t}^{GS} \leq P_{\max}^{GS} \tag{12}$$

where Eq. (2) defines the power balance constraint. Eq. (3) defines the PV output power limits. Eq. (4) defines the MT output power limits, while Eq. (5) enforces the MT ramping constraints. The ESS model is governed by Eqs. (6)-(9), which define the charge/discharge power limits, state of charge (SOC) evolution, and SOC bounds, respectively. ESS operational costs are neglected in this work. Eq. (10) enforces the exclusivity of power exchange states between each MG and the upstream grid, which ensures that purchase and sell operations cannot occur simultaneously. Eqs. (11) and (12) impose limits on the aggregated power procured from and sold to the upstream grid, respectively.

*C. MARL Formulation for Multi-MG Energy Scheduling*

In multi-MG energy scheduling, the dispatch strategies of MGs are highly interdependent, as the actions of one agent affect the outcomes for all others. This interactive decision-making process is well-suited for the Markov Game framework, which effectively models the system dynamics and aligns individual strategies with global objectives.

This work models the multi-MG energy scheduling problem as a Markov Game, formally defined by the tuple $\langle \mathcal{N}, \mathcal{S}, \mathcal{O}, \mathcal{A}, \mathcal{R}, \mathcal{P}, \gamma \rangle$. Each MG deploys an agent $i \in \mathcal{N}$ to interact with the environment, where $\mathcal{N} = \{1, ..., n\}$ is the set of agents; $s \in \mathcal{S}$ denotes global state space; $\{o_i \in \mathcal{O}_i\}$ denotes partially observable state space; $\{a_i \in \mathcal{A}_i\}$ denotes action space; $\{r_i \in \mathcal{R}_i\}$ denotes reward space; $\mathcal{P}(s' | s, a)$ denotes state transition function; $\gamma \in [0,1)$ denotes reward discount factor.

Each agent $i$ learns a decentralized policy $\pi_{\theta_i}(a_{i,t} | o_{i,t})$ parameterized by $\theta_i$. The joint policy $\pi_\theta$ aims to maximize the expected discounted cumulative return over the entire scheduling horizon:

$$\mathcal{J}(\pi_\theta | s_t) = E_{\pi_\theta} \left[ \sum_{t=1}^{T} \sum_{i=1}^{N} \gamma^t r_{i,t}(a_{i,t} \sim \pi_\theta | o_{i,t}) \right], \forall i, \forall o_i \in \mathcal{O}_i \tag{13}$$

where $\mathcal{J}$ denotes the global reward function. $r_{i,t} \in \mathcal{R}$ represents the immediate reward of agent $i$.

*1) State and observation:* The global state and the local observation at timestep $t$ are represented by Eqs. (14) and (15), respectively. The local observation is a subset of the global state accessible to each agent.

$$s_t = (o_{1,t}, ..., o_{n,t}) \tag{14}$$

$$o_{i,t} = (t, \lambda_t^{\text{in}}, P_{i,t}^{PV}, P_{i,t}^{MT}, P_{i,t}^{L}, SOC_{i,t}) \tag{15}$$

*2) Action:* The action of each agent is defined as the power outputs of its ESSs and MTs, subject to the constraints given in Eqs. (4)-(7).

$$a_{i,t} = (P_{i,t}^{ES}, P_{i,t}^{MT}) \tag{16}$$

where $P_{i,t}^{ES}$ represents the net power instruction for the ESS, with negative values indicating charging power $P_{i,t}^c$ and positive values indicating discharging power $P_{i,t}^d$. Note that trading with the grid is not an action but a consequence determined to maintain power balance after the ESS and MT actions are set [19]. If the resulting power exchange at timestep $t$ violates the limits, load shedding is enacted with a penalty.

*3) Reward Function:* The reward function is defined as the negative of the total operational cost to incentivize economically optimal policies.

$$r_{i,t} = -(\lambda_t^{MT} P_{i,t}^{MT} + \lambda_t^{\text{in}} P_{i,t}^{GB} - \lambda_t^{\text{out}} P_{i,t}^{GS} + \lambda_t^{TL} P_{i,t}^{TL}) \tag{17}$$

In the proposed framework, all agents jointly maximize a global reward. To ensure global optimality, each agent's individual value function is aligned with this collective objective [20]. However, the globally optimal policy may require some MGs to undertake economically suboptimal actions (e.g., operating high-cost gas turbines) for collective benefit. To maintain coalition stability and incentivize participation, this work adopts the Shapley Value for fair cost allocation based on marginal contributions. For any coalition $S \subseteq \mathcal{N}$, its characteristic function $Y(S)$ is defined as the minimum total utility (operational cost and load-shedding risk) under coordinated energy scheduling by its members. $Y(S)$ represents the optimal solution found by the RRL-SM framework when applied only to the agents in $S$, which is formulated in Eq. (18). Based on the Shapley Value, the fair cost allocation $\Psi_i(Y)$ for MG $i$ is calculated in Eq. (19).

$$Y(S) = \min_{\pi_S} E_{\tau \sim \pi_S} \left[ C(S) + CVaR_\alpha(S) \right] \tag{18}$$

$$\Psi_i(Y) = \sum_{S \subseteq \mathcal{N} \setminus \{i\}} \frac{|S|!(n-|S|-1)!}{n!} \cdot \left[ Y(S \cup \{i\}) - Y(S) \right] \tag{19}$$

where $S$ is a sub-coalition that does not include MG $i$. $Y(S)$ is the minimum total utility achievable by coalition $S$ without MG $i$. $CVaR_\alpha(S)$ quantifies the load shedding risk, as elaborated in Section III-A. $Y(S\cup\{i\})$ is the minimum total utility of the coalition after MG $i$ joins. $[Y(S\cup\{i\})-Y(S)]$ represents the marginal cost contribution of agent $i$ to coalition $S$.

Note that this cost allocation is a post-processing mechanism designed to incentivize MG participation in coordinated energy scheduling. This mechanism does not alter the training objective of the RRL-SM framework, which remains the maximization of the total system reward.

## III. RISK-SENSITIVE MARL

### A. Risk Formulation

To address the operational challenges stemming from the uncertainty of PV output and load demand, this paper adopts a scenario-based methodology. This methodology generates a set of representative scenarios to model the uncertainty, and then employs a coherent risk measure to assess the potential for adverse outcomes.

*1) Scenario Generation for Uncertainty*

To obtain representative operational scenarios, a large ensemble of stochastic profiles for PV output and load demand are generated via Monte Carlo simulation based on historical data. This set is subsequently reduced to a computationally tractable number using the k-means clustering algorithm [21].

$$\Omega^{PV} = \{\mu_{b,1},\ldots,\mu_{b,t},\ldots,\mu_{b,T}\}, b=1,2,\ldots B \tag{20}$$

$$\Omega^{Load} = \{\mu_{d,1},\ldots,\mu_{d,t},\ldots,\mu_{d,T}\}, d=1,2,\ldots D \tag{21}$$

$$\Pr_{\Omega^{PV}} = [p_1,\ldots,p_b,\ldots,p_B]^T, \sum_{b=1}^{B} p_b = 1 \tag{22}$$

$$\Pr_{\Omega^{Load}} = [q_1,\ldots,q_d,\ldots,q_D]^T, \sum_{d=1}^{D} q_d = 1 \tag{23}$$

where the possible operating scenarios of the PV output and load demand are represented by Eqs. (20) and (21), respectively. Eqs. (22) and (23) define the probability associated with each scenario.

*2) Risk Assessment via Conditional Value-at-Risk*

Operational risk is defined as the potential for substantial economic losses resulting from load shedding. To quantify this risk, especially the impact of high-consequence, low-probability events (i.e., tail risk), this paper employs Conditional Value-at-Risk (CVaR) [22]. CVaR measures the expected loss in the worst-case quantile of a given loss distribution. Within the proposed MARL framework, the centralized critic network uses this risk measure to formulate a risk-sensitive learning signal. This signal then guides the decentralized agents to learn coordinated, risk-averse policies. Given a confidence level $\alpha$, the $CVaR_\alpha$ discretized expression considering the source-load uncertainties is formulated as:

$$CVaR_\alpha = VaR_\alpha + \frac{1}{1-\alpha}\sum_{b=1}^{B}\sum_{d=1}^{D} p_b q_d [\vartheta_{b,d} - VaR_\alpha]^+ \tag{24}$$

$$[\vartheta_{b,d} - VaR_\alpha]^+ = \max(\vartheta_{b,d} - VaR_\alpha, 0) \tag{25}$$

$$\vartheta_{b,d} = \sigma \sum_{i=1}^{N}\sum_{t=1}^{T} P_{i,t}^{TL} \tag{26}$$

where $\vartheta_{b,d}$ denotes the total system expense under a composite scenario, which combines the $b$-th PV output scenario with probability $p_b$ and the $d$-th load scenario with probability $q_d$. $VaR_\alpha$ represents the maximum value of loss that could be incurred by a dispatch strategy.

### B. Risk-sensitive Value Factorization for cooperative MARL

The prevailing paradigm for cooperative MARL is CTDE. Within the CTDE framework, value factorization methods (VDN, QMIX, etc) have become mainstream approaches for ensuring policy coordination. These approaches rely on the Individual-Global-Max (IGM) principle, which assumes that the joint action-value function could be additively decomposed into individual utilities. However, this assumption becomes problematic in risk-sensitive settings, as common risk measures (e.g., VaR, CVaR) are inherently non-additive. Consequently, global risk objectives cannot be directly distributed among agents via simple summation or monotonic mixing. For instance, consider the $VaR_{0.5}$ metric (the median). In general, $VaR_{0.5}[Z_1+Z_2] \neq VaR_{0.5}[Z_1]+VaR_{0.5}[Z_2]$, since the median of a sum does not equal the sum of medians. This non-additivity precludes standard factorization forms from correctly propagating risk-sensitive objectives across agents.

The distributional reinforcement learning paradigm enables the direct quantification and optimization of risk by modeling the entire distribution of returns, rather than only its expectation. This capability allows for the formulation of new coordination criteria suited to risk-sensitive tasks. This work adopts a coordination principle for risk-sensitive value factorization [23]. As formally defined in Eq. (27), this principle requires that the optimal risk-sensitive joint action be equivalent to the collection of each agent's individually optimal risk-sensitive actions.

$$\arg\max_a \varphi_\alpha[Z_{\text{tot}}(\tau,a)] = (\arg\max_{a_1} \varphi_\alpha[Z_1(\tau_1,a_1)],\ldots,\arg\max_{a_n} \varphi_\alpha[Z_n(\tau_n,a_n)]) \quad (27)$$

where $Z_{\text{tot}}(\tau,a)$ denotes the joint state-action return distribution. $Z_i(\tau_i,a_i)$ is agent $i$'s individual return distribution. $\tau_i$ is the local action-observation history $(o_i^1,a_i^1,\ldots,o_i^{t-1},a_i^{t-1},o_i^t)$. $\varphi_\alpha(\cdot)$ is a risk measure (e.g., VaR, CVaR).

To satisfy the coordination requirement formally defined in Eq. (27), this paper proposes a risk-sensitive value factorization method. This method models the mapping between individual and joint return distributions using quantile functions. It provides the necessary flexibility to factorize non-additive risk measures. Since value factorization is typically applied to state-action values, its integration into a policy-based framework like multi-agent proximal policy optimization (MAPPO) requires modification [24]. Specifically, the factorization is applied to the state value distribution $Z_i(o_i)$, which is employed by the critic network in guiding the policy updates of the actors. The state return distribution is represented by its quantile function:

$$\theta_Z(o,\omega) = \inf\{z \in \mathbb{R} : \omega \leq CDF_Z(z)\}, \forall \omega \in [0,1] \quad (28)$$

where $\omega$ is a quantile level. $CDF_Z(z)$ is the cumulative distribution function (CDF). $\theta_Z(\omega)$ is simplified to $\theta(\omega)$.

The proposed method first represents each agent's return distribution $Z_i(o_i)$ with its quantile function $\theta_i(o_i,\omega_j)$. Then, a multi-head attention mechanism generates a dynamic weighting function, which is combined with the individual quantile functions to calculate the joint return distribution:

$$Z_{\text{tot}}(s) = \sum_{j=1}^{J} p_j(s,\omega_j) \delta_{\theta(s,\omega_j)} \quad (29)$$

$$\theta(s,\omega_j) = \sum_{i=1}^{N} k_i(s)\theta_i(o_i,\omega_j) \quad (30)$$

where $J$ indicates the number of quantile points. $\delta_{\theta(s,\omega_j)}$ is the Dirac delta function. $p_j(s,\omega_j)$ denotes the probability associated with each quantile point.

The attention mixer employs multi-head attention to capture inter-MG coordination patterns and dynamically adjust risk-sensitive quantile weights $k_i(s)$ [25]. $k_i(s)$ is formulated in Eq. (31). The integrated coordination context $\mathbf{c}_i$ for each microgrid is constructed by concatenating all attention heads and then projecting the result through $W^O$, as formulated in Eq. (32). Each specialized attention head $\text{head}_i^h$ computes distinct coordination patterns using linearly transformed input features through query, key, and value matrices, as defined in Eq. (33). The core attention mechanism in Eq. (34) computes similarity scores between queries and keys, followed by scaled value aggregation. The input feature sequence $X_i$ encoding temporal dependencies through GRU processing is given in Eq. (35).

$$k_i(s) = \text{softmax}(\mathbf{u}^\top \tanh(\mathbf{W}\mathbf{c}_i + \mathbf{b})) \quad (31)$$

$$\mathbf{c}_i = \text{MultiHead}_i(Q,K,V) = \text{Concat}\left[\text{head}_i^1,\ldots,\text{head}_i^H\right]W^O \quad (32)$$

$$\text{head}_i^h = \text{Attention}(X_i W_h^Q, X_i W_h^K, X_i W_h^V) \quad (33)$$

$$\text{Attention}(Q,K,V) = \text{softmax}\left(\frac{QK^\top}{\sqrt{d_k}}\right)V \quad (34)$$

$$X_i = \text{GRU}(s^{1:t}) \quad (35)$$

where, $\mathbf{u},\mathbf{W},\mathbf{b}$ are learnable parameters. For each attention head $h=1,\ldots,H$, $W_h^Q \in \mathbb{R}^{d_{\text{model}} \times d_k}$, $W_h^K \in \mathbb{R}^{d_{\text{model}} \times d_k}$, $W_h^V \in \mathbb{R}^{d_{\text{model}} \times d_k}$, and $W^O \in \mathbb{R}^{H \cdot d_k \times d_{\text{model}}}$ denote the query weight matrix, key weight matrix, value weight matrix, and output projection matrix, respectively. $d_{\text{model}}$ and $d_k = d_{\text{model}}/H$ represent the input feature dimension and key vector dimension, respectively. $\sqrt{d_k}$ denotes the square root of the key vector dimension [25].

*C. Shared Memory Mechanism*

Explicit communication becomes inefficient or intractable as the number of agents and problem dimensionality increase. Moreover, individual memory architectures are insufficient for complex inter-agent coordination. A novel shared memory mechanism is proposed to achieve scalable, implicit coordination. This mechanism enhances agent coordination efficiency by establishing a global information space. The architecture of the shared memory module is illustrated in Fig. 2.

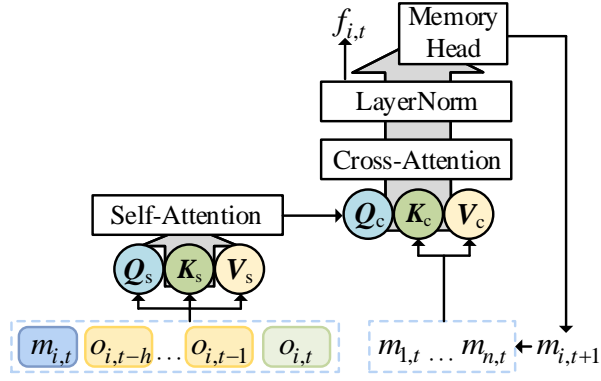

Fig. 2. Overview of the shared memory module.

*1) Private Memory Processing (Self-Attention)*

First, each agent processes its private sequence (its own memory $m_{i,t}$ and its observation history $o_{i,t-h},...,o_{i,t-1}$) using self-attention to generate a context-aware representation, $\text{SelfAttn}_{i,t}$:

$$\text{PrivateSequence}_{i,t} = [m_{i,t}, o_{i,t-h}, ..., o_{i,t-1}, o_{i,t}] \tag{36}$$

$$\boldsymbol{Q}_s, \boldsymbol{K}_s, \boldsymbol{V}_s = \text{Linear}(\text{PrivateSequence}_{i,t}) \tag{37}$$

$$\text{SelfAttn}_{i,t} = \text{softmax}\left(\frac{\boldsymbol{Q}_s \boldsymbol{K}_s^\top}{\sqrt{d_k}}\right) \boldsymbol{V}_s \tag{38}$$

where $\boldsymbol{Q}_s, \boldsymbol{K}_s, \boldsymbol{V}_s \in \mathbb{R}^{L \times d_k}$ are query, key, and value matrices derived from the private sequence. $L$ is the sequence length. $d_k$ is the dimension of each attention head. $\text{Linear}(\cdot)$ is a standard linear transformation layer. $\text{softmax}(\cdot)$ is a standard softmax function.

*2) Global Memory Space (Cross-Attention)*

Next, the shared memory space $\mathcal{M}_t = \{m_{i,t},...,m_{n,t}\}$ serves as global workspace. The agent uses its context-aware representation $\text{SelfAttn}_{i,t}$ to generate a new query $\boldsymbol{Q}_c$. It uses this query to access the global shared memory $\mathcal{M}_t$ via cross-attention [26]:

$$\boldsymbol{Q}_c = \text{Linear}(\text{SelfAttn}_{i,t}) \tag{39}$$

$$\boldsymbol{K}_c, \boldsymbol{V}_c = \text{Linear}(\mathcal{M}_t) \tag{40}$$

$$\text{CrossAttn}_{i,t} = \text{softmax}\left(\frac{\boldsymbol{Q}_c (\boldsymbol{K}_c)^\top}{\sqrt{d_k}}\right) \boldsymbol{V}_c \tag{41}$$

where $\boldsymbol{K}_c$ and $\boldsymbol{V}_c$ are derived from $\mathcal{M}_t$, enabling global state awareness without explicit communication.

*3) Memory Update and Decision Vector Generation*

Finally, the output of the cross-attention is normalized. The updated representation of the memory token $m_{i,t+1}$ and the final decision vector $f_{i,t}$ for the actor network are extracted from the resulting tensor $\mathbf{H}_{i,t}$.

$$\mathbf{H}_{i,t} = \text{LayerNorm}(\text{CrossAttn}_{i,t}) \tag{42}$$

$$m_{i,t+1} = \mathbf{W}_m \cdot \mathbf{H}_{i,t}[0,:] \tag{43}$$

$$f_{i,t} = \mathbf{H}_{i,t}[-1,:] \tag{44}$$

where $\text{LayerNorm}(\cdot)$ is the layer normalization operation. $\mathbf{W}_m$ is the memory update weight matrix.

## IV. ALGORITHM DESIGN AND IMPLEMENTATION

### A. RRL-SM Framework

To achieve an effective balance among coordination, training efficiency, and risk mitigation, this paper extends a risk-sensitive value factorization method to the MAPPO framework. The system's functionalities are structured around three components: agent-environment interaction, shared memory module, and risk-sensitive value factorization. The architecture of the proposed RRL-SM framework is illustrated in Fig. 3.

Specifically, in agent-environment interaction, each agent's actor network independently generates actions based on a processed decision vector derived from the shared memory, and subsequently interacts with the environment. The shared memory module operates as follows: initially, the shared memory core processes private data (including the agent's own memory, historical observations, and current observations) via a self-attention layer. Subsequently, each agent queries the shared memory space through a cross-attention layer, thereby achieving global state awareness. Finally, this process generates updated decision vectors for actors and updates the agents' internal memories.

In the risk-sensitive value factorization component, a global, risk-sensitive value is computed to guide policy updates. This is achieved through a sophisticated centralized critic network comprising individual quantile critic networks, a mixing network, and a risk module. Initially, to capture return uncertainty, each agent's associated quantile critic network receives a local observation and in turn outputs a discrete quantile distribution of the state-value. A multi-head attention mechanism is employed to compute dynamic correlation weights between agents. Subsequently, the mixing network utilizes these weights to combine the individual return distributions, yielding a joint state-value return distribution. Finally, a dedicated risk module computes the risk-sensitive joint state value and risk advantage functions, which are utilized to update the actor networks' parameters.

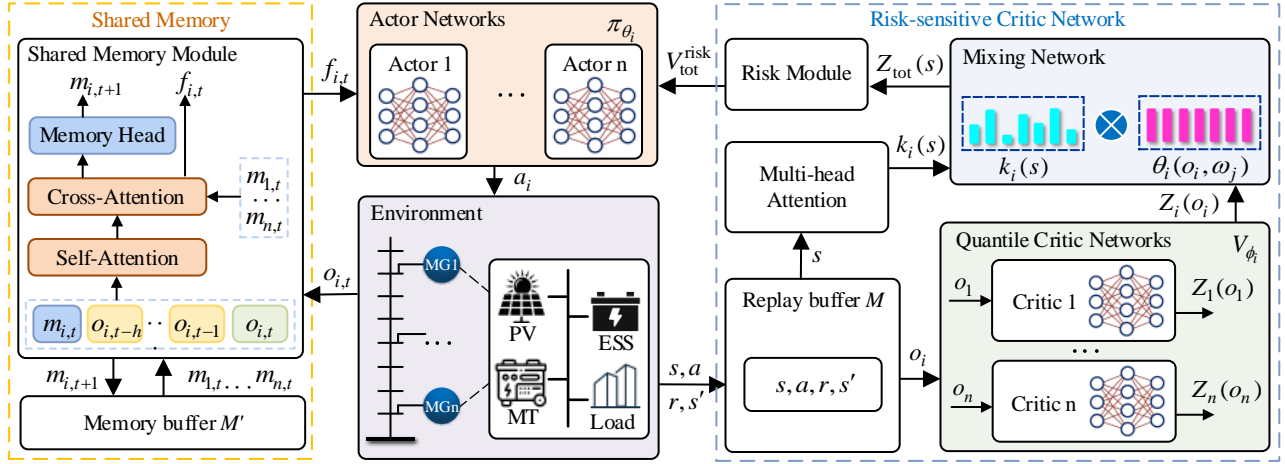

Fig. 3. Overview of the RRL-SM architecture.

### B. Risk-sensitive Critic Network

Inspired by the value-decomposition actor-critic (VDAC) framework [24], this paper proposes a risk-sensitive critic network as the core of the value factorization approach. It provides a global, risk-sensitive learning signal to guide the decentralized actor networks.

*1) Quantile Critic Network:*

To capture return uncertainty, each agent is associated with a quantile critic network. Standard critic network outputs a single expected value, which represents the average return but completely obscures the underlying distribution of potential outcomes and their associated risks. In contrast, the quantile critic network for each agent receives a local observation $o_i$ and outputs a discrete quantile distribution of the state-value $Z_i(o_i)$. This full distributional output is the essential foundation for the subsequent risk assessment performed by the Risk Module.

*2) Attention Mixer:*

The attention mixer takes the individual return distributions $\{Z_1(o_1),...,Z_n(o_n)\}$ from all agents and the global state $s$ as inputs. It employs a multi-head attention mechanism to learn a dynamic, state-dependent weighting function, which blends the quantile functions of each agent's return distribution into a joint return distribution $Z_{tot}(s)$. Since each agent's contribution to global risk varies with the environment state, the attention mechanism enables the mixer to infer complex inter-agent dependencies and adapt to varying risk contribution patterns across states.

*3) Risk Module:*

The risk module transforms the complex joint return distribution $Z_{tot}$ from the attention mixer into a single actionable risk metric for decision-making under uncertainty. It uses a coherent risk measure applied to the distribution. This calculation yields a scalar value representing the expected loss in the worst-case scenarios, thereby capturing tail risk. This scalar serves as a learning signal for decentralized actors, promoting globally coordinated and risk-averse policies.

$$V_{tot}^{risk}(s) = \varphi_\alpha [Z_{tot}(s)] \tag{45}$$

where $V_{tot}^{risk}(s)$ is the risk-sensitive joint state value, which is defined as a scalar measure of the risk-adjusted value for the global state $s$. $\varphi_\alpha(\cdot)$ is the risk measure operator. It maps the entire return distribution to a single scalar risk value, and its calculation is formally defined in Eqs. (24)-(26). $\alpha$ denotes the confidence level for the risk measure (e.g., $\alpha = 0.9$ corresponds to the worst 10% of outcomes).

The quantile critic network is trained to minimize the discrepancy between the predicted and target joint return distributions. The target distribution is calculated using the Bellman equation. This objective is achieved by minimizing the following quantile regression loss. Eq. (46) defines the overall loss function for the quantile critic network. The loss is calculated by summing the errors across $J$ different quantile points of the distribution. By minimizing this loss, the quantile critic network learns to approximate the shape of the true return distribution [27]. Eq. (48) defines an asymmetric loss for a single quantile. Its design penalizes overestimations and underestimations differently. For a given quantile level $v_j$, if the prediction is too low ($u > 0$), the error is weighted by $v_j$. If the prediction is too high ($u < 0$), the error is weighted by $1 - v_j$. This asymmetry forces the network to learn the specific location of each quantile, thereby capturing the overall shape of the distribution. Eq.(49) defines the Huber loss, which is used within the quantile regression loss to improve training stability. It behaves like a squared-error ($\ell 2$) loss for small errors and a linear-error ($\ell 1$) loss for large errors.

$$\mathcal{L}_{\text{Critic}}(\phi_i) = \sum_{j=1}^{J} E_{(Z, Z_{\text{target}}) \sim M} \left[ \rho_\kappa^{v_j} \left( TD(z_j) - z_j(s; \phi) \right) \right] \tag{46}$$

$$TD(z_j) = r_t + \gamma z_j(s_{t+1}) \tag{47}$$

$$\rho_\kappa^{v_j}(u) = \left| v_j - \chi(u < 0) \right| \frac{\mathcal{L}_\kappa(u)}{\kappa} \tag{48}$$

$$\mathcal{L}_\kappa(u) = \begin{cases} 0.5 u^2 & \text{if } |u| \leq \kappa \\ \kappa(|u| - 0.5\kappa) & \text{otherwise} \end{cases} \tag{49}$$

where $\mathcal{L}_{\text{Critic}}(\phi_i)$ is the loss function for the quantile critic network parameterized by $\phi_i$. $TD(z_j)$ is the target quantile value for the $j$-th quantile, usually computed using a target network and the Bellman equation. $z_j(s; \phi)$ is the predicted quantile value for the $j$-th quantile at state $s$. $u$ is the error term between the target and predicted quantile values. $v_j$ is the j-th quantile level, a value in the range (0, 1) (e.g., 0.1, 0.5, 0.9). $\chi(u < 0)$ is an indicator function, which equals 1 if u < 0 and 0 otherwise. $\kappa$ defines the boundary between small and large errors [23].

### C. Actor Network

Each agent possesses a decentralized actor network that takes a decision vector $f_{i,t}$ and outputs a probability distribution over its actions $a_{i,t} \sim \pi_i(\cdot | f_{i,t})$. The actor network is updated by maximizing the clipped surrogate objective of PPO [27]:

$$\mathcal{L}_{\text{actor}}(\theta_i) = E_t \left[ \min \left( \zeta_t(\theta) \hat{A}_t, \text{clip}(\zeta_t(\theta), 1-\varepsilon, 1+\varepsilon) \hat{A}_t \right) \right] \tag{50}$$

$$\zeta_t(\theta) = \frac{\pi_\theta(a_t | s_t)}{\pi_{\theta_{old}}(a_t | s_t)} \tag{51}$$

$$\hat{A}_t = \sum_{l=0}^{T-t-1} (\gamma \lambda)^l \delta_{t+l} \tag{52}$$

$$\delta_t = r_t + \gamma V_{\text{tot}}^{\text{risk}}(s_{t+1}) - V_{\text{tot}}^{\text{risk}}(s_t) \tag{53}$$

where $\zeta_t(\theta)$ is the probability ratio between the new and old policies; clip(·) is a function that constrains the probability ratio $\zeta_t(\theta)$ to stay within the range $[1-\varepsilon, 1+\varepsilon]$; $\varepsilon$ is the clipping hyperparameter. The advantage function $\hat{A}_t$ is calculated with general advantage estimation (GAE), which uses the joint risk state-value from the centralized critic network as its baseline. $\lambda$ is the GAE parameter, which controls the bias-variance trade-off; $\delta_{t+l}$ is the temporal difference (TD) error at a future timestep $t+l$ [28].

Finally, the pseudo-code of RRL-SM for training process is shown as Algorithm 1, and the proof of local convergence for RRL-SM is provided in the appendix.

---

**Algorithm 1** RRL-SM for training process

1: Initialize actor networks $\theta_i$, quantile critic networks $\phi_i$, and shared memory parameters
2: Initialize replay buffer $M$
3: Set learning rates and clip factor $\varepsilon$
4: **for** episode $= 1, 2, ...$ **do**
5:     Initialize global state $s_0$ and local observations $o_{i,0}$
6:     Initialize shared memory space $\{m_{i,0}, ..., m_{n,0}\}$
7:     **for** $t = 0, 1, ..., T-1$ **do**
8:         Compute cross-attention with shared memory using (36)-(41)
9:         Update memory using (43): $m_{i,t+1}$
10:        Generate decision vector using (44): $f_{i,t}$
11:        For each agent $i$, select action $a_{i,t} \sim \pi_{\theta_i}(\cdot | f_{i,t})$
12:        Get rewards $r_t$ and next state $s_{t+1}$
13:        Store transition $(s_t, a_{i,t}, r_{i,t}, s_{t+1})$ in buffer $M$
14:        For each agent $i$, get individual return distribution from its quantile critic network: $Z_i(o_{i,t})$
15:        Aggregate individual distributions into a joint return distribution using (29)-(30): $Z_{\text{tot}}(s_t)$
16:        Compute the risk-sensitive joint state value from the joint return distribution using (45): $V_{\text{tot}}^{\text{risk}}(s_t)$
17:     **end for**
18:     Sample batch of trajectories from buffer $M$
19:     Compute advantages $\hat{A}_t$ using GAE with $V_{\text{tot}}^{\text{risk}}(s_t)$
20:     Update quantile critic networks $\phi_i$ by minimizing $\mathcal{L}_{\text{critic}}(\phi_i)$ using (46)-(49)

21:     Update actor networks $\theta_i$ by maximizing $\mathcal{L}_{\text{actor}}(\theta_i)$ using (50)-(53)
22: **end for**

## V. CASE STUDY

### A. Experimental Setup

The simulation environment is a multi-MG system consisting of three MGs interconnected with the upstream grid. The scheduling horizon is set to 168 hours (one week) with a one-hour time step. To evaluate the generalization capability of the proposed method, three months of historical data were partitioned into a training set (first 10 weeks) and a testing set (remaining 2 weeks) [29]. Using independent random seeds, a large ensemble of 1000 stochastic scenarios are generated from the training set via Monte Carlo simulation to capture the full spectrum of uncertainties in PV output and load demand. For computational tractability during the training and evaluation phases, this large set is then reduced to 20 representative training scenarios and 20 independent test scenarios (detailed in Section II-C). This ensures mutually exclusive scenario sets for training and evaluation. During each training episode, a scenario is randomly sampled from the 20 representative training scenarios to update the policy. The final performance is evaluated on the held-out test scenarios to ensure generalization. The cost coefficients for the MT are adopted from [5], and other key parameters are listed in Table I.

TABLE I. SPECIFIC CONFIGURATION PARAMETERS

| Symbol | Value | Symbol | Value |
|---|---|---|---|
| $\eta^c / \eta^d$ | 0.98/0.98 | $B$ | 20 |
| $\lambda_t^{\text{out}}$ | $0.5 \cdot \lambda_t^{\text{in}}$ | $D$ | 20 |
| $E_{i,\max}$ | 400(kWh) | $p_b / q_d$ | 0.05/0.05 |
| $P_{i,\max}^c / P_{i,\max}^d$ | 200(kW) | $T$ | 168 |
| $\lambda_i^{\text{MT}}$ | 1.01(CNY/kWh) | $\alpha$ | 90% |
| $\lambda_i^{\text{TL}}$ | 10(CNY/kWh) | $H$ | 4 |
| $P_{i,\max}^{\text{PV}} / P_{i,\min}^{\text{PV}}$ | 350 /0(kW) | $p_j(s, \omega_j)$ | 1/32 |
| $P_{i,\max}^{\text{MT}} / P_{i,\min}^{\text{MT}}$ | 300 /0(kW) | $J$ | 32 |
| $R_i^{\text{up}} / R_i^{\text{down}}$ | 100 /100(kW/h) | $n$ | 3 |
| $P_{\max}^{\text{GB}} / P_{\max}^{\text{GS}}$ | 900(kW) | $\gamma$ | 0.99 |
| Actor Learning Rate | 1e-4 | $\varepsilon$ | 0.2 |
| Critic Learning Rate | 5e-4 | $\lambda$ | 0.95 |
| Hidden Layer Size | 64 | $\kappa$ | 10 |

| Symbol | Time Slot | (CNY /kWh) |
|---|---|---|
| $\lambda_t^{\text{in}}$ | 00:00-6:00, 21:00-24:00 | 0.423 |
| | 6:00-7:00, 10:00-12:00,15:00-17:00 | 0.775 |
| | 7:00-10:00, 12:00-15:00, 17:00-21:00 | 1.189 |

All computations were accelerated on an NVIDIA GeForce RTX 5090 D GPU (32 GB VRAM), with supporting hardware including an AMD Ryzen 9 9950X processor and 128 GB RAM. Model development and training were implemented in Python using PyTorch.

### B. Convergence Performance

To evaluate the effectiveness of the RRL-SM algorithm, this paper compares it against three baseline algorithms: QMIX [7], MAPPO [27], and a variant of MAPPO that incorporates risk-sensitive value factorization (denoted as R-MAPPO). Fig. 4 illustrates the cumulative reward curves for the RRL-SM, R-MAPPO, MAPPO, and QMIX algorithms, respectively. Note that a scenario is randomly selected in each episode, and a fixed random seed ensures all algorithms face the same sequence of scenarios. Consequently, the observed fluctuations in the curves primarily stem from the inherent variations in cumulative rewards across different scenarios. Algorithm performance is evaluated based on the overall reward level and its stability over an extended training horizon.

Compared to QMIX, both RRL-SM and MAPPO exhibit superior convergence stability. In contrast, the reward curves for QMIX and R-MAPPO show significant volatility after approximately 500 training episodes. The volatility observed in QMIX likely stems from its difficulty in effectively coordinating agents within a complex environment with non-stationary rewards. In contrast, the fluctuations in R-MAPPO are attributed to the real-time coordination challenges inherent in its decentralized decision-making process. The final cumulative reward of R-MAPPO is slightly lower than that of MAPPO. This is because the algorithm's risk-averse mechanism guides agents toward a more conservative policy, highlighting an inherent trade-off. Although the adopted policy may not be the most economically optimal, it substantially enhances operational reliability by mitigating risk. RRL-SM improves upon R-MAPPO by incorporating a shared memory module, which significantly enhances the online coordination capabilities among agents. Consequently, RRL-SM achieves a higher cumulative reward than R-MAPPO.

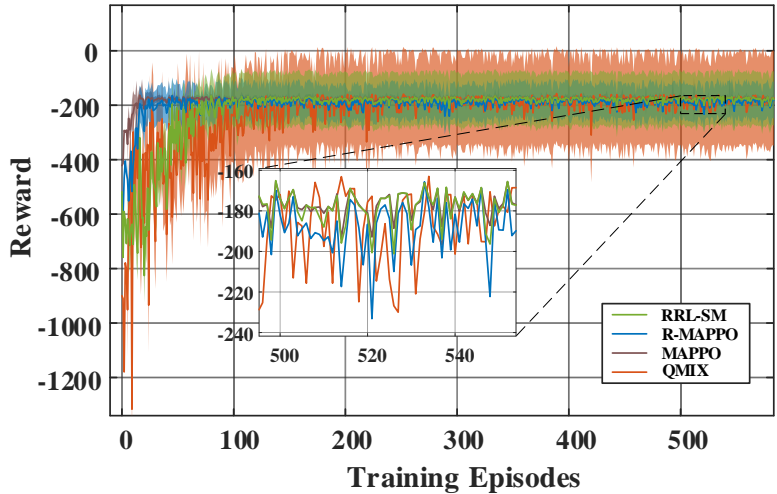

Fig. 4. Comparison of cumulative rewards.

## C. Analysis of Scheduling Strategy

As shown in Figs. 5-7, the scheduling strategies of the three microgrids collectively demonstrate the core feature of coordinated risk-aversion in the proposed framework. Specifically: MG 1, with low daytime load, charges its energy storage at midday and increases MT output in the afternoon to prepare for evening risks. MG 2 utilizes its ESS for energy time-shifting. This involves charging the ESS during periods of high PV generation and discharging it during sharp declines in PV output or at times of high load demand. Although MG 3 has relatively stable load throughout the day, it faces the risk of sudden drops in PV output and therefore relies on its ESS to compensate for the resulting power deficit.

These strategies all embed a proactive risk mitigation mechanism. The algorithm anticipates that evening load peak could cause grid transactions to exceed their limits, load shedding or risking penalties. To mitigate this risk, the algorithm takes preemptive actions much earlier in the day. It strategically increases the MT output and pre-charges the ESS even when immediate costs are not minimal. This builds up sufficient reserve capacity to ensure the evening peak demand is met without violating grid constraints, thus clearly demonstrating a risk-averse, coordinated scheduling behavior.

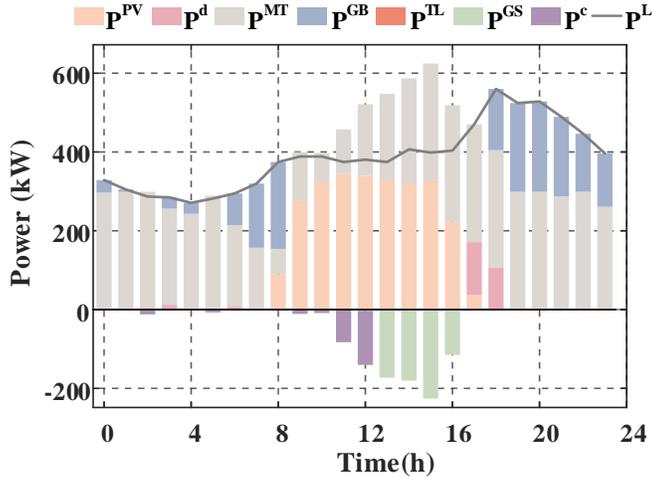

Fig. 5. Energy scheduling results of MG 1 on a typical day.

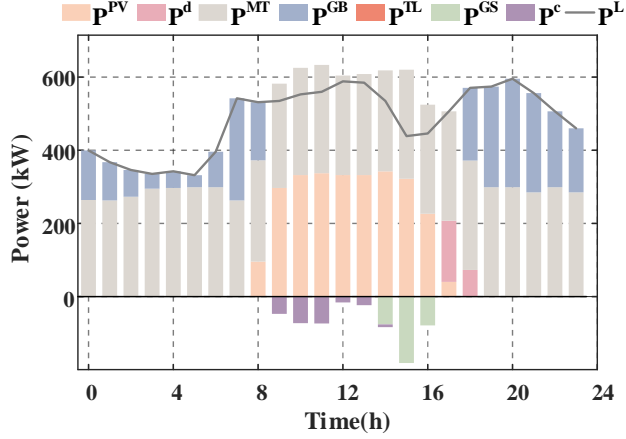

Fig. 6. Energy scheduling results of MG 2 on a typical day.

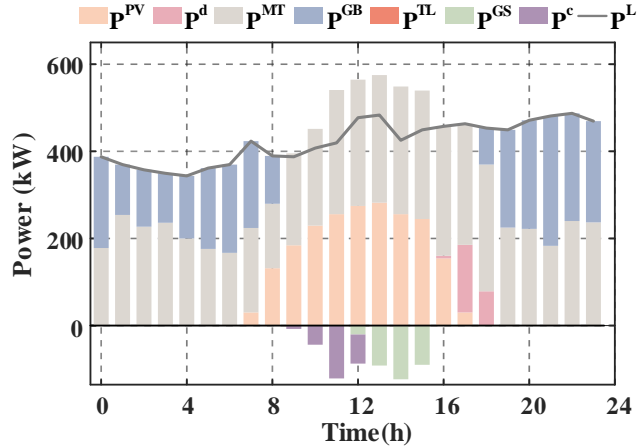

Fig. 7. Energy scheduling results of MG 3 on a typical day.

### D. *Economic and Risk Assessment*

Table II presents a performance comparison of the four algorithms. The results reveal that although MAPPO and QMIX achieve lower operational costs, this comes at the expense of a significantly higher risk of load shedding. In contrast, R-MAPPO cuts the load-shedding risk (measured as CVaR) by 85.3% compared to MAPPO, with only a 2.7% loss in economic efficiency. This result effectively validates the effectiveness of its risk-control mechanism in ensuring system operational reliability. RRL-SM reduces load-shedding risk by 84.5% compared to MAPPO, while increasing operational cost by only 1.8%. Furthermore, for the proposed RRL-SM method, we compare its performance under different confidence levels, which demonstrates its ability to balance cost and risk. Building upon this foundation, the proposed RRL-SM method leverages its shared memory module to simultaneously reduce operational costs and ensure robust risk control.

TABLE II. COMPARISON OF COST AND LOAD-SHEDDING RISK.

| Method | Confidence Level | Total Cost (CNY) | Unit Cost (CNY/kWh) | Risk (kWh) |
|---|---|---|---|---|
| RRL-SM | 0.9 | 176677 | 0.7378 | **72.74** |
|  | 0.5 | 175953 | 0.7348 | 162.78 |
| R-MAPPO | 0.9 | 178235 | 0.7443 | **68.93** |
| MAPPO | - | 173537 | 0.7247 | 470.51 |
| QMIX | - | 174817 | 0.7301 | 687.17 |

To validate the effectiveness of the risk-sensitive value factorization method, this study compares the decision-making performance of the R-MAPPO and MAPPO algorithms in a one-week electricity trading scenario involving three MGs and the upstream grid. Fig. 8 and 9 present the trading results under the respective algorithms. On the third day, characterized by higher load demand and a significant increase in electricity purchases from the upstream grid, R-MAPPO leveraged its risk-sensitive mechanism to identify potential load-shedding risk. By proactively increasing gas turbine output and energy storage charging, it achieved preemptive mitigation of limit violation risks, thereby substantially reducing the probability of load shedding. In contrast,

MAPPO exhibited transaction limit violations under the same conditions. Despite a higher coordination penalty coefficient, it failed to suppress these violations entirely.

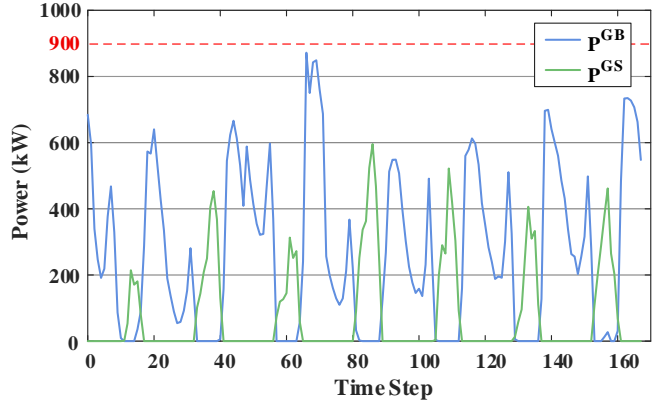

Fig.8. Weekly grid trading power under R-MAPPO.

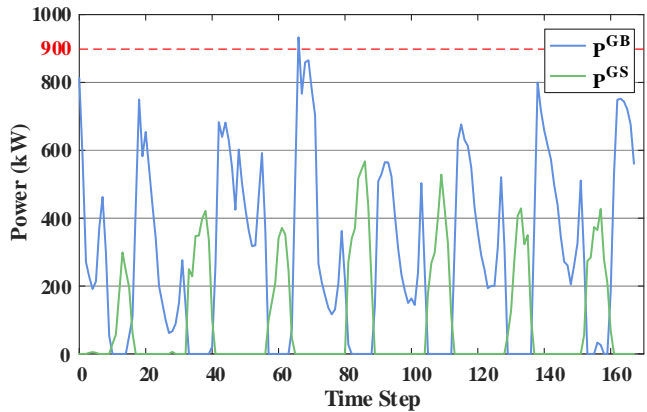

Fig.9. Weekly grid trading power under MAPPO.

*E. Effect of Confidence Level on Dispatch Results*

To evaluate the impact of decision-makers' risk preferences on scheduling strategies, a comparative analysis was conducted by setting different confidence levels. By definition, a higher confidence level reflects a lower risk preference, while a lower confidence level indicates a higher risk tolerance.

Fig. 10 illustrates the dispatch strategy under a high confidence level ($\alpha = 0.9$). This strategy focuses on avoiding the worst 10% of extreme scenarios, showing typical risk-averse characteristics. The results indicate that during daytime peak solar generation periods, the system uses energy storage charging and selling electricity to the grid to absorb the solar power. This pre-charging of storage acts as an effective forward risk-hedging measure. It shifts daytime energy to the evening peak hours, ensuring load supply after the sudden drop in solar generation and reducing the risk of load shedding caused by transaction limit violations with the grid.

In contrast, Fig. 11 shows the strategy under a low confidence level ($\alpha = 0.5$). To reduce immediate generation costs, this strategy reduces gas turbine output during periods 15–17, compensating for the shortfall with energy storage and power purchases from the grid. Although this approach lowers short-term costs, it neglects the ramp-rate constraints of gas turbines. As a result, the system cannot respond quickly when the load peak arrives at period 18, introducing significant load-shedding risk. This comparison clearly reveals the trade-off between immediate cost savings and long-term system security under different risk preferences. This conclusion is further supported by the system operation costs and load shedding risks listed in Table II.

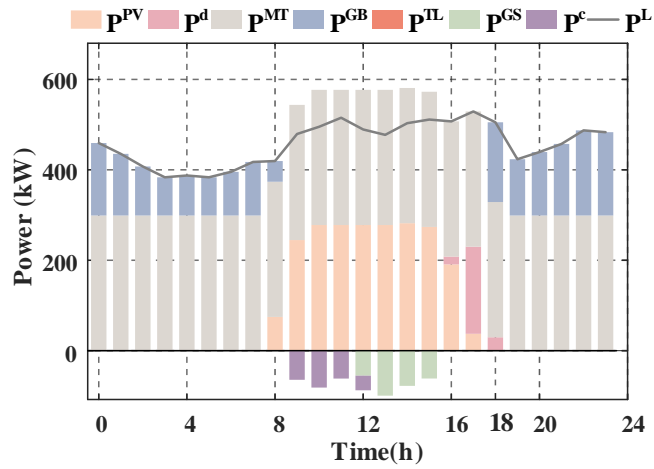

Fig.10. Energy scheduling results of MG 2 on a typical day under a high confidence level.

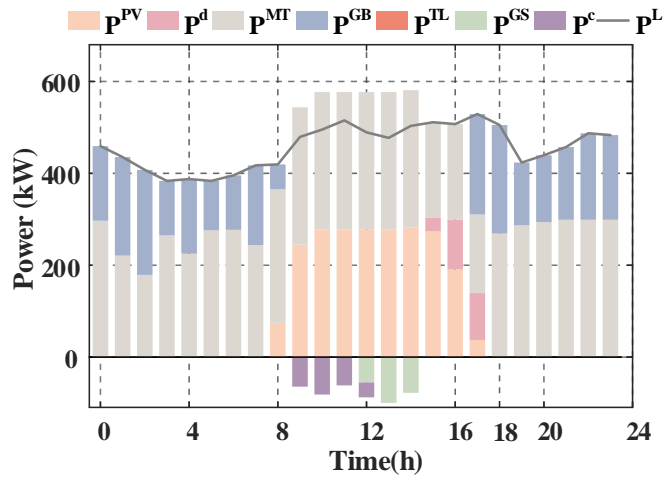

Fig.11. Energy scheduling results of MG 2 on a typical day under a low confidence level.

## VI. Conclusion

This paper proposes the RRL-SM framework to address the challenge of risk-coordinated scheduling for multi-MG systems under uncertainty. This algorithm integrates a risk-sensitive value factorization method and a novel shared memory mechanism. Case studies show that the R-MAPPO algorithm effectively reduces the load-shedding risk compared to the MAPPO and QMIX algorithms, at the expense of a slight reduction in economic efficiency. The proposed RRL-SM framework successfully overcomes this trade-off by using its shared memory module. This allows the algorithm to achieve superior economic performance compared to R-MAPPO and maintain robust risk control. The effectiveness of the RRL-SM framework in balancing operational reliability and economic efficiency is validated.

## Acknowledgment

This work was supported by the Guangxi Natural Science Foundation under Grant 2025GXNSFBA069263; and the Guangxi Science and Technology Program under Grant No.AB24010255.

## References

[1] Marqusee J, Becker W, Ericson S. Resilience and economics of microgrids with PV, battery storage, and networked diesel generators. Advances in Applied Energy 2021;3:100049. https://doi.org/10.1016/j.adapen.2021.100049.
[2] Daneshvar M, Mohammadi-Ivatloo B, Zare K. A Fair Risk-Averse Stochastic Transactive Energy Model for 100% Renewable Multi-Microgrids in the Modern Power and Gas Incorporated Network. IEEE Trans Smart Grid 2023;14:1933–45. https://doi.org/10.1109/TSG.2022.3218255.
[3] Siqin T, He S, Hu B, Fan X. Shared energy storage-multi-microgrid operation strategy based on multi-stage robust optimization. Journal of Energy Storage 2024;97:112785. https://doi.org/10.1016/j.est.2024.112785.
[4] Wang Y, Qiu D, Sun M, Strbac G, Gao Z. Secure energy management of multi-energy microgrid: A physical-informed safe reinforcement learning approach. Applied Energy 2023;335:120759. https://doi.org/10.1016/j.apenergy.2023.120759.
[5] Herding R, Ross E, Jones WR, Endler E, Charitopoulos VM, Papageorgiou LG. Risk-aware microgrid operation and participation in the day-ahead electricity market. Advances in Applied Energy 2024;15:100180. https://doi.org/10.1016/j.adapen.2024.100180.
[6] Xu X, Xu K, Zeng Z, Tang J, He Y, Shi G, et al. Collaborative optimization of multi-energy multi-microgrid system: A hierarchical trust-region multi-agent reinforcement learning approach. Applied Energy 2024;375:123923. https://doi.org/10.1016/j.apenergy.2024.123923.

[7] Zhang M, Tong W, Zhu G, Xu X, Wu EQ. SQIX: QMIX Algorithm Activated by General Softmax Operator for Cooperative Multiagent Reinforcement Learning. IEEE Transactions on Systems, Man, and Cybernetics: Systems 2024;54:6550–60. https://doi.org/10.1109/TSMC.2024.3370186.
[8] Zhu Z, Gao X, Bu S, Chan KW, Zhou B, Xia S. Cooperative Dispatch of Renewable-Penetrated Microgrids Alliances Using Risk-Sensitive Reinforcement Learning. IEEE Transactions on Sustainable Energy 2024;15:2194–208. https://doi.org/10.1109/TSTE.2024.3406590.
[9] Salyani P, Zare K, Javani N, Boynuegri AR. Risk-based scheduling of multi-energy microgrids with Power-to-X technology under a multi-objective framework. Sustainable Cities and Society 2025;122:106245. https://doi.org/10.1016/j.scs.2025.106245.
[10] Datta J, Das D. Risk-constrained Energy Management of multi-energy multi-microgrids system with integrated demand response using hybrid risk assessment approach. Energy 2025;331:136638. https://doi.org/10.1016/j.energy.2025.136638.
[11] Liu Z, Zhao H, Liu G, Liang G, Zhao J, Qiu J. Risk-Sensitive Mobile Battery Energy Storage System Control With Deep Reinforcement Learning and Hybrid Risk Estimation Method. IEEE Transactions on Smart Grid 2024;15:4143–58. https://doi.org/10.1109/TSG.2024.3358838.
[12] Skrynnik A, Andreychuk A, Nesterova M, Yakovlev K, Panov A. Learn to Follow: Decentralized Lifelong Multi-Agent Pathfinding via Planning and Learning. AAAI 2024;38:17541–9. https://doi.org/10.1609/aaai.v38i16.29704.
[13] Hu Y, Fu J, Wen G. Graph Soft Actor–Critic Reinforcement Learning for Large-Scale Distributed Multirobot Coordination. IEEE Trans Neural Netw Learning Syst 2025;36:665–76. https://doi.org/10.1109/TNNLS.2023.3329530.
[14] Wang Y, Xiang B, Huang S, Sartoretti G. SCRIMP: Scalable Communication for Reinforcement- and Imitation-Learning-Based Multi-Agent Pathfinding. 2023 IEEE/RSJ International Conference on Intelligent Robots and Systems (IROS), Detroit, MI, USA: IEEE; 2023, p. 9301–8. https://doi.org/10.1109/IROS55552.2023.10342305.
[15] Yang Y, Chen G, Wang W, Hao X, Hao J, Heng PA. Transformer-based Working Memory for Multiagent Reinforcement Learning with Action Parsing. vol. 35, 2022, p. 34874–86.
[16] Bulatov A, Kuratov Y, Burtsev MS. Recurrent Memory Transformer. vol. 36, 2022, p. 1049–5258.
[17] Shi H, He Z, Hwang K-S. Adaptive path planning for wafer second probing via an attention-based hierarchical reinforcement learning framework with shared memory. Information Sciences 2025;710:122089. https://doi.org/10.1016/j.ins.2025.122089.
[18] Li Z, Wu L, Xu Y. Risk-Averse Coordinated Operation of a Multi-Energy Microgrid Considering Voltage/Var Control and Thermal Flow: An Adaptive Stochastic Approach. IEEE Trans Smart Grid 2021;12:3914–27. https://doi.org/10.1109/TSG.2021.3080312.
[19] Shengren H, Vergara PP, Salazar Duque EM, Palensky P. Optimal energy system scheduling using a constraint-aware reinforcement learning algorithm. International Journal of Electrical Power & Energy Systems 2023;152:109230. https://doi.org/10.1016/j.ijepes.2023.109230.
[20] Qiu D, Wang Y, Wang J, Zhang N, Strbac G, Kang C. Resilience-Oriented Coordination of Networked Microgrids: A Shapley Q-Value Learning Approach. IEEE Trans Power Syst 2024;39:3401–16. https://doi.org/10.1109/TPWRS.2023.3276827.
[21] Xiaoyu Wang, Ying Han, Luoyi Li, Jinxuan Wang, Weirong Chen, Wenjie Shen. CVaR Quantitative Uncertainty-Based Optimal Dispatch for Flexible Traction Power Supply System. IEEE Transactions on Transportation Electrification 2024;10:1900–10. https://doi.org/10.1109/TTE.2023.3272358.
[22] Zhang Q, Leng S, Ma X, Liu Q, Wang X, Liang B, et al. CVaR-Constrained Policy Optimization for Safe Reinforcement Learning. IEEE Trans Neural Netw Learning Syst 2025;36:830–41. https://doi.org/10.1109/TNNLS.2023.3331304.
[23] Shen S, Ma C, Li C, Liu W, Fu Y, Mei S, et al. RiskQ: Risk-sensitive Multi-Agent Reinforcement Learning Value Factorization. Advances in Neural Information Processing Systems, vol. 36, Curran Associates, Inc.; 2023, p. 34791–34825.
[24] Su J, Adams S, Beling P. Value-Decomposition Multi-Agent Actor-Critics. AAAI 2021;35:11352–60. https://doi.org/10.1609/aaai.v35i13.17353.
[25] Huang Y, Fan F, Huang C, Yang H, Gu M. MA-DG: Learning Features of Sequences in Different Dimensions for Min-Entropy Evaluation via 2D-CNN and Multi-Head Self-Attention. IEEE TransInformForensic Secur 2024;19:7879–94. https://doi.org/10.1109/TIFS.2024.3447242.
[26] Huo G, Zhang Y, Gao J, Wang B, Hu Y, Yin B. CaEGCN: Cross-Attention Fusion Based Enhanced Graph Convolutional Network for Clustering. IEEE Trans Knowl Data Eng 2023;35:3471–83. https://doi.org/10.1109/TKDE.2021.3125020.
[27] Hassan SS, Park YM, Tun YK, Saad W, Han Z, Hong CS. SpaceRIS: LEO Satellite Coverage Maximization in 6G Sub-THz Networks by MAPPO DRL and Whale Optimization. IEEE J Select Areas Commun 2024;42:1262–78. https://doi.org/10.1109/JSAC.2024.3369665.
[28] Kang H, Chang X, Mišić J, Mišić VB, Fan J, Liu Y. Cooperative UAV Resource Allocation and Task Offloading in Hierarchical Aerial Computing Systems: A MAPPO-Based Approach. IEEE Internet Things J 2023;10:10497–509. https://doi.org/10.1109/JIOT.2023.3240173.
[29] Ordoudis C, Pinson P, Morales JM, Zugno M. An Updated Version of the IEEE RTS 24-Bus System for Electricity Market and Power System Operation Studies. Technical University of Denmark; 2016.

APPENDIX A

**Proof.** The proof of local convergence for the RRL-SM algorithm is based on the policy gradient theorem. The proof consists of two parts: (1) deriving the policy gradient for the multi-agent case, and (2) showing that the variance-reducing baseline does not introduce bias.

The policy gradient is given by:

$$\nabla_\theta J(\theta) = E_\pi \left[ \sum_{i=1}^{N} \nabla_\theta \log \pi_i(a_i \mid o_i) \cdot A(s,a) \right] \tag{A.1}$$

where $A(s,a) = Q(s,a) - V_{\text{tot}}(s)$ is the advantage function.

We decompose the policy gradient into the action-value term $G_Q$ and the baseline term $G_V$. With the joint policy factorization $\pi(a \mid s) = \sum_{i=1}^{N} \pi_i(a_i \mid o_i)$ in decentralized execution, $G_V$ is rewritten as:

$$\begin{aligned} G_V &= -E_\pi \left[ \sum_{i=1}^{N} \nabla_\theta \log \pi_i(a_i \mid o_i) \cdot V_{\text{tot}}(s) \right] \\ &= -E_\pi \left[ \nabla_\theta \log \prod_{i=1}^{N} \pi_i(a_i \mid o_i) \cdot V_{\text{tot}}(s) \right] \\ &= -E_\pi \left[ \nabla_\theta \log \pi(a \mid s) \cdot V_{\text{tot}}(s) \right] \end{aligned} \tag{A.2}$$

where, the distribution $E_\pi$ is over states and actions induced by the joint policy $\pi$. Let $d^{\pi(s)}$ denote the stationary state distribution under policy $\pi$. Under the CTDE framework, $V_{tot}(s)$ depends solely on the state $s$ and is independent of the specific action selection. Consequently, the baseline term vanishes:

$$\begin{aligned} G_V &= -\sum_s d^{\pi(s)} \sum_a \nabla_\theta \log \pi(a|s) \cdot V_{tot}(s) \\ &= -\sum_s d^{\pi(s)} V_{tot}(s) \nabla_\theta \sum_a \log \pi(a|s) \\ &= -\sum_s d^{\pi(s)} V_{tot}(s) \nabla_\theta 1 = 0 \end{aligned} \quad (A.3)$$

The remainder of the gradient is given by:

$$\begin{aligned} \nabla_\theta J(\theta) &= E_\pi \left[ \sum_{i=1}^N \nabla_\theta \log \pi_i(a_i|o_i) \cdot Q(s,a) \right] \\ &= E_\pi \left[ \nabla_\theta \log \prod_{i=1}^N \pi_i(a_i|o_i) \cdot Q(s,a) \right] \\ &= E_\pi \left[ \nabla_\theta \log \pi(a|s) \cdot Q(s,a) \right] \end{aligned} \quad (A.4)$$

Under standard regularity conditions, an iterative update following the unbiased gradient converges to a local optimum of the expected return [24]. Therefore, the RRL-SM algorithm is guaranteed to converge to a locally optimal joint policy.